\documentclass[%
 aip,
 jmp,%
 amsmath,amssymb,
 reprint,%
]{revtex4-1}

\usepackage{graphicx}
\usepackage{dcolumn}
\usepackage{bm}

\usepackage[cal=boondox]{mathalfa}

\usepackage{amsmath}
\usepackage{relsize}

\DeclareMathAlphabet\mathbfcal{OMS}{cmsy}{b}{n}

\begin{document}

\preprint{AIP/123-QED}

\title[The Method of Formal Series...]{The Method of Formal Series: Applications to Nonlinear Beam Dynamics and Invariants of Motion}

\author{Stephan I. Tzenov}
\email{tzenov@jinr.ru, \newline {\it Alternative electronic mail}: stephan.tzenov@eli-np.ro}

\affiliation{Zhangjiang Laboratory, 99 Haike Rd., Pudong New District, Shanghai, China} 

\affiliation{Joint Institute for Nuclear Research, 6 Joliot-Curie Street, Dubna, Moscow Region, Russian Federation, 141980}




\date{\today}

\begin{abstract}
A novel technique to determine invariant curves in nonlinear beam dynamics based on the method of formal series has been developed. It is first shown how the solution of the Hamilton equations of motion describing nonlinear betatron oscillations in the presence of a single sextupole can be represented in a nonperturbative form. Further, the solution of the Hamilton-Jacobi equation is obtained in a closed symbolic form as a ratio of two series in the perturbation parameter (and the nonlinear action invariant), rather than a conventional power series according to canonical perturbation theory. It is well behaved even for large values of the perturbation parameter close to strong structural resonances. The relationship between existing invariant curves and the so-called scattering orbits in classical scattering theory has been revealed. 
\end{abstract}

\maketitle

%

\section{\label{sec:intro}Introduction}

"Yet another method in nonlinear dynamics on the horizon..." - thus, some potential readers of this article would be skeptical. Such a reaction is quite understandable, since the literature abounds in publications devoted to various approaches to the study of nonlinear dynamical systems. Without claiming completeness, we will try to provide a short review on the available methods, especially those that have found application in the nonlinear dynamics of charged particle beams. At the same time, we shall indicate where in this multitude the method of formal series developed here stands. 

The origins of the theory of dynamical systems are assumed to be the end of the nineteenth century with Henri Poincare \cite{Poincare}, who is considered the father of modern nonlinear dynamics. Another prominent giant from the early years is undoubtedly Alexander Lyapunov, who formulated the basis of the theory of stability, \cite{Lyapunov} which is important in all critical instances of nonlinear dynamics marking the transition between different qualitative behaviors. They are followed by a whole galaxy of remarkable researchers mathematicians and physicists like Birkhoff, Kolmogorov, Arnold, Moser and many others, who have left a deep trace behind. It is not possible to give the deserved attention to all of them, and that is not our goal here.

The most popular and widespread analytical asymptotic approaches to solve nonlinear equations of motion are the different variants of mathematical techniques united by a common feature and known under the umbrella name as perturbative methods. These are the most extensively used approximate methods in studying nonlinear dynamics models arising in many branches of physics, engineering and in particular the dynamics of charged particle beams. A major limitation of these methods is the restriction related to the existence of a small parameter which makes the solution valid for weakly nonlinear systems mostly, and in addition, for relatively short time intervals marking the short term dynamics of nonlinear systems. 

One of the first versions of the perturbative approach in Hamiltonian mechanics is the canonical theory of perturbations; an accessible overview of which can be found in the book of Lichtenberg and Lieberman \cite{Lichtenberg}. Whether based on classical mechanics, like the Birkhoff-Gustavson method \cite{Gustavson} and alternatively the Lie transform technique \cite{Dragt}, or the quantum mechanics counterpart, like the Van-Vleck procedure \cite{Kemble}, the purpose of canonical perturbation theory is always to cast the Hamiltonian of the original system in terms of, as complete as possible, a set of classical invariants of motion (action variables), or good quantum numbers, respectively. The transformed Hamiltonian, whose dynamics can usually be analyzed in great detail thanks to the conserved quantities, then provides the keys for understanding the principal properties of the initial system. A crucial problem of the canonical perturbation theory occurs even in the case of nonlinear motion in a single degree of freedom. Nevertheless such systems are known to be completely integrable, a naive perturbation expansion in the amplitude of the oscillation alone leads to the appearance of secular terms (the so-called small denominators, or terms as a minimum linearly increasing with time) and subsequent divergence of the series. What to speak in the case of many degrees of freedom, where the dynamics are much richer in features and sources of secular behavior. The first step in remedying the deficiency of small denominators has been made by Lindstedt \cite{Lindstedt} and Poincare \cite{PoincareB}, who proposed to vary the amplitude and the frequency of nonlinear oscillations together. Let us note that the Lindstedt-Poincare method is closely related to the method of multiple scales \cite{Kevorkian}, where the dynamics are separated into rapidly oscillating and slow motion, each of them characterized by its intrinsic time (frequency) scales. A variation of the canonical perturbation theory, which prescribes how to remove the resonant secular terms, is known as the secular perturbation theory \cite{Chirikov}. There exist another version of the secular perturbation theory, which in certain cases can be applied to remove an entire set of resonances, also known as the method of Dunnett-Laing-Taylor \cite{Dunnett}. 

An important achievement and a milestone in the theory of perturbations applied to Hamiltonian systems is the introduction of Lie transformations by Hori \cite{Hori}. The Lie transformation method was further improved by Deprit, \cite{Deprit} who managed to obtain expressions in a closed form for the $n$th-order term of an expansion of the transformation in a power series in a formal small parameter. An equivalent formulation of the Lie's algebraic Hori-Deprit method, which is more known and widely used in the modern physics of accelerators and charged particle beams, is the Dragt-Finn approach \cite{Dragt}. 

In the majority of practically important cases an analytical solution to the equations of motion is a hopeless exercise, so as the necessity of employing numerical methods arises. Either way, all numerical methods for solving differential equations involve discretization schemes, so it is natural to pose the question about the possibility of substituting the Hamilton’s equations of motion with mapping. In contemporary literature, there is an abundance of articles devoted to the application of nonlinear (Lie or Taylor) maps in beam dynamics, making it impossible to mention them all. With the awareness of possible reproaches for bias or personal preference, we will constraint ourselves here to mentioning only some of these references \cite{DragtAIP,DragtBOOK,Forest,Berz,Turchetti}, which became classical in the field. For a more complete picture, one may also consult the literature cited therein. 

As already noted above, the main disadvantage of all perturbative methods is the presence of resonant small denominators which makes the solution, generally speaking, valid only in an asymptotic sense. The method of multiple scales (respectively the Lindstedt-Poincare method), as well as the renormalization group method \cite{tzenovBOOK,Nambu} do a little better. In both approaches, the motion is divided into rapid oscillations (fast varying phases) and relatively slow dynamics of the corresponding amplitudes on longer time scales. The renormalization group method gives a prescription on how to remove the divergent secular terms and as a result to obtain the so-called envelope equations describing the formation of various patterns and coherent structures. Among the nonperturbative methods in nonlinear mechanics, it is worth noting the interesting method proposed by Delamotte \cite{Delamotte}, where the nonlinearity is appropriately approximated by an external force acting on a linear oscillator. 

The mathematical details concerning the method of formal series have been considered in the most precise and exhaustive way in Ref. \citenum{DuboisViolette}. It is there shown that this method can be applied to solve a number of nonlinear equations in mathematical physics of a special class (as well as equations that can be transformed to such class). Further, this approach has been applied to study the peculiarities in the dynamics of an anharmonic oscillator in classical mechanics \cite{DuboisVioletteMech}, as well as to build a functional formulation of quantum field theory \cite{DuboisVioletteNC}. A further elaboration has been carried out in Refs. \citenum{TzenovArt,tzenovBOOK}, where the method has been adapted to study nonlinear dynamics in accelerators and storage rings. The basic idea of the method of formal series is to represent the solution of a nonlinear equation as a ratio of two formal Volterra series in powers of the perturbation parameter, rather than a conventional power series, prescribed by classical perturbation theory. It turns out that the behaviour of the ratio between the two series is good enough even for large values of the coupling parameter. This fact should not be so surprising, if the solution by the method of formal series is interpreted as a nonlinear generalization of the solution of a simple linear system of equations by a ratio of two determinants. Moreover, the denominator is uniquely defined, which is not the case with Pade approximations \cite{Pade} and other nonperturbative tools, where one should fix one or more coefficients in the series expansion of the denominator, taking into account initial conditions. Nevertheless, the solution of the initial problem obtained by the method of formal series is given in symbolic form (in a symbolic form the general solution by the Lie transformation technique is presented as well), one is granted the possibility to compute both the numerator and the denominator to any desired order with respect to the perturbation parameter. The method of formal series has been applied successfully to solve nonlinear partial differential equations when studying various collective wave processes in nonneutral plasmas \cite{TzenovELI1,TzenovELI2}. 

The structure of the article is as follows. For the completeness and self-content of the exposition, Section \ref{sec:basics} is devoted to a brief introduction of the method of formal series as applied to solve a system of nonlinear algebraic equations. In this simplest case, the solution is represented by an ordered divergence operator acting on a known function, which in a sense is inverse to the translation operator. The demonstration of the basic idea of our approach on a simple algebraic example has been carried out for a methodological purpose, which makes straightforward its generalization in the space of real valued functions. The latter is described in detail in Section \ref{sec:funcgen}. Since the solution of the nonlinear integral equation (of Volterra or Fredholm type) is given in a symbolic form, further clarification on the way it should be applied has been provided. In Section \ref{sec:solham} the application of the method of formal series to the solution of the Hamilton’s equations of motion describing the betatron oscillations in a plane transversal to the particle orbit has been discussed. A specific choice of a single sextupole perturbing the linear machine lattice has been studied in greater detail. In Section \ref{sec:hamiljacobi} the Hamilton-Jacobi equation is solved by the method of formal series. It has been assumed that the new transformed Hamiltonian depends only on the new action variable, which implies that the sought-for generating function describes the invariant curves. This specificity is not a restriction; the new Hamiltonian can be taken in an arbitrary normal in the sense of Birkhoff form. Numerical results demonstrating its application on a linear accelerator lattice perturbed by a single sextupole, or an octupole have been also presented. Finally, in Section \ref{sec:concluding} our conclusions and future perspectives are sketched out.  

\section{\label{sec:basics}Brief Mathematical Foundations of the Method of Formal Series for Solving of Nonlinear Algebraic Equations}

Since the method of formal series is little known in the field of nonlinear dynamics, it is worth starting with its brief description. To this end, we shall follow closely Ref. \citenum{tzenovBOOK}, while rigorous mathematical details can be found in Ref. \citenum{DuboisViolette}. To begin with, let us demonstrate on the simplest case of a system of nonlinear algebraic equations, how the method of formal series works. Let us consider the following system 
\begin{equation}
{\bf x} + \epsilon {\bf g} {\left( {\bf x} \right)} = {\boldsymbol{\xi}}, \label{NLinSystem}
\end{equation}
where the right-hand-side ${\boldsymbol{\xi}} = {\left( \xi_1, \xi_2, \dots, \xi_N \right)}$ represents a $N$-dimensional vector playing the role of $N$ independent variables, ${\bf x} = {\left( x_1, x_2, \dots, x_N \right)}$ is the unknown vector-variable and ${\bf g} {\left( {\bf x} \right)} = {\left[ g_1 {\left( {\bf x} \right)}, g_2 {\left( {\bf x} \right)}, \dots, g_N {\left( {\bf x} \right)} \right]}$ is a given vector-function of the unknown variables $x_1, x_2, \dots, x_N$. Clearly, the solution to Eq. (\ref{NLinSystem}) will be expressed in terms of some vector-function of the independent variables and of the (not necessarily small) parameter $\epsilon$ as well. Thus, we have 
\begin{equation}
{\bf x} = {\bf f} {\left( {\boldsymbol{\xi}} \right)}, \qquad \quad {\rm so \; that} \qquad \quad {\boldsymbol{\xi}} = {\bf f}^{-1} {\left( {\bf x} \right)}. \nonumber
\end{equation}
Plugging the second expression of the above equation into the right-hand-side of Eq. (\ref{NLinSystem}), we can rewrite the latter as follows 
\begin{equation}
{\bf x} + \epsilon {\bf g} {\left( {\bf x} \right)} = {\bf f}^{-1} {\left( {\bf x} \right)}. \label{NLinSyst}
\end{equation}
Inverting Eq. (\ref{NLinSyst}), we write it in alternative form 
\begin{equation}
{\bf x} = {\bf f} {\left( {\bf x} + \epsilon {\bf g} {\left( {\bf x} \right)} \right)}. \label{NLinSysAlt}
\end{equation}
Let us formally replace ${\bf x}$ with ${\boldsymbol{\xi}}$ in Eq. (\ref{NLinSysAlt}), so that 
\begin{equation}
{\boldsymbol{\xi}} = {\bf f} {\left( {\boldsymbol{\xi}} + \epsilon {\bf g} {\left( {\boldsymbol{\xi}} \right)} \right)}. \label{NLinSysReplace}
\end{equation}
This latter equation implies that the sought-for vector-function ${\bf f} {\left( {\boldsymbol{\xi}} \right)}$ is just the inverse of the translation operator by a vector-argument, that is precisely the given vector-function ${\bf g} {\left( {\boldsymbol{\xi}} \right)}$. In a symbolic form, Eq. (\ref{NLinSysReplace}) can be written as follows 
\begin{equation}
{\widehat{\boldsymbol{\mathfrak{L}}}} \exp {\left[ \epsilon {\bf g} {\left( {\boldsymbol{\xi}} \right)} \cdot {\boldsymbol{\nabla}}_{\xi} \right]} \bullet {\bf f} {\left( {\boldsymbol{\xi}} \right)} = {\boldsymbol{\xi}}. \label{NLinSysSymb}
\end{equation}
The symbol ${\widehat{\boldsymbol{\mathfrak{L}}}}$ implies that after the operator $\exp {\left[ \dots \right]}$ on the left-hand-side of Eq. (\ref{NLinSysSymb}) has been developed in power series, the gradient operators ${\boldsymbol{\nabla}}_{\xi}$ should be shifted to the right of the multipliers (functions) ${\bf g} {\left( {\boldsymbol{\xi}} \right)}$, so that their action extends on the function ${\bf f} {\left( {\boldsymbol{\xi}} \right)}$ only. The notation "$\bullet {\bf f}$" to the right of the operator ${\widehat{\boldsymbol{\mathfrak{L}}}} \exp {\left( \dots \right)}$ signifies that the latter is an operator acting on a given function ${\bf f}$.

Let us now introduce the Gamma-operator 
\begin{equation}
{\widehat{\boldsymbol{\Gamma}}} \exp {\left[ - \epsilon {\boldsymbol{\nabla}}_{\xi} \cdot {\bf g} {\left( {\boldsymbol{\xi}} \right)} \right]} \bullet , \label{GammaOper}
\end{equation}
which is inverse in some sense to the generalized translation operator ${\widehat{\boldsymbol{\mathfrak{L}}}} \exp {\left[ \epsilon {\bf g} {\left( {\boldsymbol{\xi}} \right)} \cdot {\boldsymbol{\nabla}}_{\xi} \right]} \bullet$ defined above. The symbol ${\widehat{\boldsymbol{\Gamma}}}$ implies that the differentiation acts on the function ${\bf f} {\left( {\boldsymbol{\xi}} \right)}$ (which should be added to the right), as well as on the function ${\bf g} {\left( {\boldsymbol{\xi}} \right)}$. All the gradient operators present in the series expansion of the Gamma-operator defined by Eq. (\ref{GammaOper}) must be shifted to the left, so that
their action spreads on the function ${\bf g} {\left( {\boldsymbol{\xi}} \right)}$ as well. The sense in which the Gamma-operator is the inverse to the generalized translational operator is embodied in the identity 
\begin{eqnarray}
{\widehat{\boldsymbol{\Gamma}}} \exp {\left[ - \epsilon {\boldsymbol{\nabla}}_{\xi} \cdot {\bf g} {\left( {\boldsymbol{\xi}} \right)} \right]} \bullet {\widehat{\boldsymbol{\mathfrak{L}}}} \exp {\left[ \epsilon {\bf g} {\left( {\boldsymbol{\xi}} \right)} \cdot {\boldsymbol{\nabla}}_{\xi} \right]} \bullet {\bf f} {\left( {\boldsymbol{\xi}} \right)} \nonumber 
\\ 
= {\bf f} {\left( {\boldsymbol{\xi}} \right)} {\widehat{\boldsymbol{\Gamma}}} \exp {\left[ - \epsilon {\boldsymbol{\nabla}}_{\xi} \cdot {\bf g} {\left( {\boldsymbol{\xi}} \right)} \right]} \bullet {\bf 1}. \label{BasicIdentity}
\end{eqnarray}
The proof of the above identity in the one-dimensional case is presented in Appendix \ref{sec:appendixA}; the generalization to the $N$-dimensional case is straightforward. Applying the Gamma-operator on both sides of Eq. (\ref{NLinSysSymb}) and solving for ${\bf f} {\left( {\boldsymbol{\xi}} \right)}$ with due account of Eq. (\ref{BasicIdentity}), we finally arrive at 
\begin{equation}
{\bf x} = {\bf f} {\left( {\boldsymbol{\xi}} \right)} = {\frac {{\widehat{\boldsymbol{\Gamma}}} \exp {\left[ - \epsilon {\boldsymbol{\nabla}}_{\xi} \cdot {\bf g} {\left( {\boldsymbol{\xi}} \right)} \right]} \bullet {\boldsymbol{\xi}}} {{\widehat{\boldsymbol{\Gamma}}} \exp {\left[ - \epsilon {\boldsymbol{\nabla}}_{\xi} \cdot {\bf g} {\left( {\boldsymbol{\xi}} \right)} \right]} \bullet {\bf 1}}}. \label{BasSol}
\end{equation}

It is worthwhile to note that the solution (\ref{BasSol}) to our initial Eq. (\ref{NLinSystem}) considered as a function of $\epsilon$ is not a conventional power series, rather than a ratio of two series. Let us recall that if the vector-function ${\bf g} {\left( {\bf x} \right)}$ is linear ${\left( {\bf g} {\left( {\bf x} \right)} = {\widehat{\bf A}} \cdot {\bf x}, \; {\rm where} \; {\widehat{\bf A}} \; {\rm is \; a} \; N \times N \; {\rm matrix} \right)}$ one arrives at a similar result. It is well-known that according to Kramer's rule the solution of a linear system of equations is expressed as a ratio of two polynomials of order $N - 1$ and $N$ respectively, rather than a single polynomial in $\epsilon$. In this sense, the solution (\ref{BasSol}) can be considered as a generalization of the determinant Kramer's rule in the nonlinear case. 

As an elementary example, let us consider the simplest example, where $g {\left( x \right)} = x$. Evaluating the numerator and the denominator of the right-hand-side of Eq. (\ref{BasSol}), we have 
\begin{eqnarray}
{\widehat{\boldsymbol{\Gamma}}} \exp {\left( - \epsilon \partial_{\xi} \xi \right)} \bullet \xi = \xi {\left( 1 - 2 \epsilon + 3 \epsilon^2 - 4 \epsilon^3 + \dots \right)} \nonumber 
\\ 
= {\frac {\xi} {{\left( 1 + \epsilon \right)}^2}}
, \nonumber
\end{eqnarray}
\begin{eqnarray}
{\widehat{\boldsymbol{\Gamma}}} \exp {\left( - \epsilon \partial_{\xi} \xi \right)} \bullet 1 = 1 - \epsilon + \epsilon^2 - \epsilon^3 + \dots \nonumber 
\\ 
= {\frac {1} {{\left( 1 + \epsilon \right)}}}
, \nonumber
\end{eqnarray}
respectively. This recovers the trivial solution of the linear algebraic equation. Similarly, one can recover Kramer's formula for the solution of a linear system of two equations with two unknowns, where ${\bf g} {\left( {\bf x} \right)} = {\left( a x_2, b x_1 \right)}$ and ${\boldsymbol{\xi}} = {\left( \xi_1, \xi_2 \right)}$  This is left to the interested reader as an exercise.



\section{\label{sec:funcgen}Functional Generalization of the Method of Formal Series} 

In order to more easily perceive the transition from linear vector spaces to the functional variant of the method of formal series, let us start with a brief introduction and reminder of some basic facts from the theory of formal Volterra series \cite{DuboisViolette,DuboisVioletteMech,DuboisVioletteNC}. In our exposition here, we will closely follow Ref. \citenum{tzenovBOOK}. For more details the interested reader is referred to Ref. \citenum{DunnBOOK} with a sufficiently accessible level of exposition. 

Let us denote by ${\boldsymbol{\mathfrak{E}}}^{\bf R}$ the space of all real valued functions $\varphi {\left( x \right)}$ varying in the interval ${\left( a, b \right)}$. For the sake of simplicity, let us consider ${\boldsymbol{\mathfrak{E}}}^{\bf R}$ a linear functional space. Assume that $\Phi_n {\left( x_1, x_2, \dots, x_n \right)}$ for $n = 0, 1, 2, \dots$ are given functions of the indicated arguments $x_i$ $\left( i = 1, 2, \dots, n \right)$, which vary also in the interval ${\left( a, b \right)}$. Without loss of generality, the functions $\Phi_n {\left( x_1, x_2, \dots, x_n \right)}$ can be taken to be symmetric in their arguments. Further, we define the functionals 
\begin{eqnarray}
P_n {\left[ \varphi \right]} = \int \limits_a^b {\rm d} x_1 \int \limits_a^b {\rm d} x_2 \dots \int \limits_a^b {\rm d} x_n \Phi_n {\left( x_1, x_2, \dots, x_n \right)} \nonumber 
\\ 
\times \varphi {\left( x_1 \right)} \varphi {\left( x_2 \right)} \dots \varphi {\left( x_n \right)}, \label{Functional}
\end{eqnarray}
with kernels $\Phi_n$, such that the integral (\ref{Functional}) exists for $\forall \varphi {\left( x \right)} \in {\boldsymbol{\mathfrak{E}}}^{\bf R}$. The series 
\begin{equation}
P {\left[ \varphi \right]} = {\mathlarger{\sum\limits_{n=0}^{\infty}}} P_n {\left[ \varphi \right]}, \label{Volterra}
\end{equation}
is called {\it formal Volterra series}. Clearly, every term and therefore the whole series (\ref{Volterra}) is a functional of $\varphi {\left( x \right)} \in {\boldsymbol{\mathfrak{E}}}^{\bf R}$. Convergence of the sum on the right-hand-side of Eq. (\ref{Volterra}) is not necessarily supposed. Thus, $P {\left[ \varphi \right]}$ is at all said a symbol -- an infinite series of numbers (asymptotic series) for which only the partial sums are well-defined. One can also introduce 
\begin{equation}
P {\left[ t, \varphi \right]} = {\mathlarger{\sum\limits_{n=0}^{\infty}}} t^n P_n {\left[ \varphi \right]}, \label{VoltFunc}
\end{equation}
which is an infinitely differentiable function with respect to the variable $t$ with
\begin{equation}
P {\left[ t, \varphi \right]} - {\mathlarger{\sum\limits_{m=0}^{N}}} t^m P_m {\left[ \varphi \right]} = O {\left( t^N \right)}. \nonumber
\end{equation}
It is possible to define a formal Volterra series resulting in a function, rather than a simple number
\begin{equation}
\Psi {\left[ x, \varphi \right]} = {\mathlarger{\sum\limits_{n=0}^{\infty}}} \Psi_n {\left[ x, \varphi \right]}, \label{VoltFunct}
\end{equation}
where 
\begin{eqnarray}
\Psi_n {\left[ x, \varphi \right]} = \int \limits_a^b {\rm d} x_1 \int \limits_a^b {\rm d} x_2 \dots \int \limits_a^b {\rm d} x_n \nonumber 
\\ 
\times G_n {\left( x; x_1, x_2, \dots, x_n \right)} \varphi {\left( x_1 \right)} \varphi {\left( x_2 \right)} \dots \varphi {\left( x_n \right)}. \label{FunFunctio}
\end{eqnarray}
Note that the kernels $G_n {\left( x; x_1, x_2, \dots, x_n \right)}$ depend on one additional argument $x$ not involved in the integration [compare with Eq. (\ref{Functional})].

One can perform all the algebraic manipulations with formal series depending on the same function $\varphi {\left( x  \right)}$. It is straightforward to introduce the sum, the product and the ratio of two formal series with a result being a formal series as well \cite{DuboisViolette}. The series
\begin{equation}
\varphi {\left[ x, \Psi \right]} = {\mathlarger{\sum\limits_{n=0}^{\infty}}} \varphi_n {\left[ x, \Psi \right]}, \label{VoltInverse}
\end{equation}
where 
\begin{eqnarray}
\varphi_n {\left[ x, \Psi \right]} = \int \limits_a^b {\rm d} x_1 \int \limits_a^b {\rm d} x_2 \dots \int \limits_a^b {\rm d} x_n \nonumber 
\\ 
\times H_n {\left( x; x_1, x_2, \dots, x_n \right)} \Psi {\left( x_1 \right)} \Psi {\left( x_2 \right)} \dots \Psi {\left( x_n \right)}, \label{FunFuncInv}
\end{eqnarray}
is said to be the inverse of the series (\ref{VoltFunct}), if being substituted back into Eq. (\ref{VoltFunct}) satisfies it identically.

A number of equations in many branches of physics and accelerator theory belong to the class of nonlinear integral equations of Volterra or Fredholm \cite{WhitBOOK} 
\begin{equation}
u {\left( x \right)} + \epsilon G {\left[ x, u \right]} = v {\left( x \right)}, \label{VoltFred}
\end{equation}
or can be transformed to such equations. Here $x$ is a real independent variable ${\left( a < x < b \right)}$, $v {\left( x \right)} \in {\boldsymbol{\mathfrak{E}}}^{\bf R}$ is a given function of the independent variable $x$ and $u {\left( x \right)}$ is the unknown function, supposed to belong to the linear functional space ${\boldsymbol{\mathfrak{E}}}^{\bf R}$. Further $G {\left[ x, u \right]}$ is an integral operator, generally nonlinear, acting on the unknown function $u {\left( x \right)}$ and transforming it to another function also belonging to ${\boldsymbol{\mathfrak{E}}}^{\bf R}$. We assume also
that the operator $G {\left[ x, u \right]}$ has a formal series expansion in $u {\left( x \right)}$ of the type (\ref{VoltFunct}) and finally, $\epsilon$ is a parameter not necessarily small.

The generalization of the expression (\ref{BasSol}) derived in Section \ref{sec:basics} expected to be suited to represent the solution to the integral equation (\ref{VoltFred}) is almost straightforward. Formally, one ought to replace the discrete vector-component index with a new continuous variable $x \in {\left( a, b \right)}$. In other words, the components $x_i$ and $\xi_i$ for ${\left( i = 1, 2, \dots, N \right)}$ become now the continuous functions $u {\left( x \right)}$ and $v {\left( x \right)}$ respectively, while the integral operator $G {\left[ x, u \right]}$ stands for the vector-function ${\bf g} {\left( {\bf x} \right)}$. The differential operator ${\boldsymbol{\nabla}}_{\xi}$ should be replaced by the functional derivative $\delta / \delta v {\left( x \right)}$, so that the expression (\ref{BasSol}) transforms to 
\begin{equation}
u {\left( x \right)} = F {\left[ x, v \right]} = {\dfrac {{\widehat{\boldsymbol{\Gamma}}} \exp {\left\{ - \epsilon {\mathlarger{\int}} {\rm d} y {\dfrac {\delta} {\delta v {\left( y \right)}}} G {\left[ y, v \right]} \right\}} \bullet v {\left( x \right)}} {{\widehat{\boldsymbol{\Gamma}}} \exp {\left\{ - \epsilon {\mathlarger{\int}} {\rm d} y {\dfrac {\delta} {\delta v {\left( y \right)}}} G {\left[ y, v \right]} \right\}} \bullet {\bf 1}}}, \label{BasSolut}
\end{equation}
respectively. It is important to emphasize that the above Eq. (\ref{BasSolut}) can be obtained directly \cite{DuboisViolette} by appropriate repetition of all the steps described in Section \ref{sec:basics}, starting from Eq. (\ref{NLinSystem}) up to Eq. (\ref{BasSol}), and utilizing the Taylor expansion for functionals 
\begin{equation}
F {\left[ y {\left( x \right)} + \varphi {\left( x \right)} \right]} = {\mathlarger{\sum\limits_{n=0}^{\infty}}} {\frac {1} {n \, !}} {\left[ y {\left( x \right)} {\frac {\delta} {\delta \varphi {\left( x \right)}}} \right]}^n F {\left[ \varphi  \right]}, \label{TaylorFunc}
\end{equation} 
proposed by Volterra, where
\begin{eqnarray}
{\left[ y {\left( x \right)} {\frac {\delta} {\delta \varphi {\left( x \right)}}} \right]}^n = \int {\rm d} x_1 \int {\rm d} x_2 \dots \nonumber 
\\ 
\times \int {\rm d} x_n y {\left( x_1 \right)} y {\left( x_2 \right)} \dots y {\left( x_n \right)} {\frac {\delta^n} {\delta \varphi {\left( x_1 \right)} \dots \delta \varphi {\left( x_n \right)}}}. \label{TaylPower}
\end{eqnarray} 

Since the solution (\ref{BasSolut}) is given in a symbolic form, it is necessary to clarify the way it should be applied. The first step
is to develop the exponential $\exp{\left\{ \dots \right\}}$ in the numerator, as well as in the denominator in a power series in $\epsilon$ (equivalently, in the functional $G$). The expression
\begin{equation}
{\left\{ {\mathlarger{\int}} {\rm d} y {\dfrac {\delta} {\delta v {\left( y \right)}}} G {\left[ y, v \right]} \right\}}^n, \nonumber
\end{equation} 
entering the coefficient of the $n$-th order term with respect to the formal parameter $\epsilon$ has to be represented in expanded form as a $n$-fold integral, which means that one must introduce different integration variables. According to the definition of the ${\widehat{\boldsymbol{\Gamma}}}$-symbol all functional derivatives should be shifted to the left of $G {\left[ y, v \right]}$, namely 
\begin{eqnarray}
{\widehat{\boldsymbol{\Gamma}}} {\left\{ {\mathlarger{\int}} {\rm d} y {\dfrac {\delta} {\delta v {\left( y \right)}}} G {\left[ y, v {\left( y \right)} \right]} \right\}}^n \nonumber 
\\ 
= \int {\rm d} y_1 \dots \int {\rm d} y_n {\frac {\delta^n G {\left[ y_1, v {\left( y_1 \right)} \right]} \dots G {\left[ y_n, v {\left( y_n \right)} \right]}} {\delta v {\left( y_1 \right)} \dots \delta v {\left( y_n \right)} }}. \label{ExpMeaning}
\end{eqnarray} 

It is worth to note that the argument $y$ in the operator $G {\left[ y, v {\left( y \right)} \right]}$ implies that the argument of the new function obtained as a result of the action of $G$ on $v {\left( \dots \right)}$ must be $y$, rather than the function $v$ enters $G$ with the same argument $y$ as in the functional derivative $\delta / \delta v {\left( y \right)}$. The next step is to substitute the representation (\ref{VoltFunct}) and (\ref{FunFunctio}) for each operator $G {\left[ y_i, v {\left( y_i \right)} \right]}$ encountered on the right-hand-side of Eq. (\ref{ExpMeaning}) and then to insert the symbols $\delta / \delta v {\left( y_i \right)}$ under the integrals coming from the representation (\ref{VoltFunct}) and (\ref{FunFunctio}). The function $v {\left( x \right)}$ in the numerator of Eq. (\ref{BasSolut}) on which the entire operator acts must be placed under the integral as well. Further, the products of all integrals thus obtained have to be represented as a multiple integral of total multiplicity, equal to the sum of integral multiplicities with which each of them enters the corresponding product. As a consequence we obtain two expressions in which all the arguments, except for the current argument $x$ are integration variables, which, in principle after performing all the integrations, should not be present in the final result. As far as the functional differentiations are concerned, one has to apply $\delta / \delta v {\left( y_i \right)}$ on a product of $v$-functions with different arguments, for instance $v {\left( z_l \right)} v {\left( z_2 \right)} \dots v {\left( z_n \right)}$. This gives 
\begin{eqnarray}
{\frac {\delta} {\delta v {\left( y_i \right)}}} v {\left( z_1 \right)} v {\left( z_2 \right)} \dots v {\left( z_m \right)} \nonumber 
\\ 
= {\mathlarger{\sum\limits_{k=1}^{m}}} v {\left( z_1 \right)} \dots v {\left( z_{k-1} \right)} \delta {\left( z_k - y_i \right)} v {\left( z_{k+1} \right)} \dots v {\left( z_m \right)}. \label{FuncDeriv}
\end{eqnarray} 
After the performance of all manipulations described above, we end up with integrals of known functions, which are products of $v$-functions and coefficients from the formal series representation of $G {\left[ y, v \right]}$. Therefore, the problem of finding the unknown function $u {\left( x \right)}$ is reduced to computing the ratio of two series with terms that are combinations of integrals with increasing multiplex. In principle it should be possible to handle these integrals.

It is interesting and important to study the possibility of extending the method of formal series to vector-function spaces. For that purpose it is necessary to define the functional version of the gradient (divergence) operator in the $N$-dimensional vector function space \cite{DuboisViolette,TzenovArt,TzenovELI1,TzenovELI2}. Suppose that instead of Eq. (\ref{VoltFred}), we have 
\begin{equation}
{\bf u} {\left( {\bf x} \right)} + \epsilon {\bf G} {\left[ {\bf x}, {\bf u} \right]} = {\bf v} {\left( {\bf x} \right)}, \label{VoltFredVec}
\end{equation}
where ${\bf u} {\left( {\bf x} \right)}$ and ${\bf v} {\left( {\bf x} \right)}$ are vector functions depending on the vector argument ${\bf x}$, and ${\bf G} {\left[ {\bf x}, {\bf u} \right]}$ is a vector functional with components of the form given by the obvious vector generalization of Eqs. (\ref{VoltFunct}) and (\ref{FunFunctio}). Then, the vector variant of Eq. (\ref{BasSolut}) is almost straightforward to express as 
\begin{equation}
{\bf u} {\left( {\bf x} \right)} = {\dfrac {{\widehat{\boldsymbol{\Gamma}}} \exp {\left\{ - \epsilon {\mathlarger{\int}} {\rm d}^N {\bf y} {\dfrac {\vec{\delta}} {\delta {\bf v} {\left( {\bf y} \right)}}} \cdot {\bf G} {\left[ {\bf y}, {\bf v} \right]} \right\}} \bullet {\bf v} {\left( {\bf x} \right)}} {{\widehat{\boldsymbol{\Gamma}}} \exp {\left\{ - \epsilon {\mathlarger{\int}} {\rm d}^N {\bf y} {\dfrac {\vec{\delta}} {\delta {\bf v} {\left( {\bf y} \right)}}} \cdot {\bf G} {\left[ {\bf y}, {\bf v} \right]} \right\}} \bullet {\bf 1}}}. \label{BasSolVec}
\end{equation}

It is worth to note that the symbolic solution expressed by Eq. (\ref{BasSolut}), or Eq. (\ref{BasSolVec}) being considered as a function of the formal parameter $\epsilon$ is not a conventional single power series, rather than a ratio of two series. It is very similar to the solution of Fredholm linear integral equations obtained by the N/D method \cite{WhitBOOK}.

\section{\label{sec:solham}Solution of the Hamilton's Equations of Motion Using the Method of Formal Series} 

In this Section we shall discuss the application of the method of formal series to the solution of the Hamilton's equations of motion describing the betatron oscillations in a plane transversal to the particle orbit. The corresponding two-degrees-of-freedom Hamiltonian can be written as 
\begin{eqnarray}
H {\left( X, P_x, Z, P_z; \theta \right)} = {\frac {{\dot{\chi}}_x {\left( \theta \right)}} {2}} {\left( P_x^2 + X^2 \right)} \nonumber 
\\ 
+ {\frac {{\dot{\chi}}_z {\left( \theta \right)}} {2}} {\left( P_z^2 + Z^2 \right)} + V {\left( X, Z; \theta \right)}, \label{HamilNorm}
\end{eqnarray}
where ${\left( X, P_x, Z, P_z \right)}$ are the normalized transverse phase-space coordinates and $\theta$ is the independent azimuthal variable matching the machine circumference. Moreover, ${\dot{\chi}}_{x,z} = R / \beta_{x,z}$ are the derivatives of the corresponding phase advances, where $R$ is the mean machine radius, and $\beta_{x,z}$ are the well-known Twiss beta-functions. The potential function $V {\left( X, Z; \theta \right)}$ is a sum of homogeneous polynomials of order three (corresponding to sextupole magnets) and higher order, characterizing static magnetic fields of higher multipolarity, if such are present. Following Ref. \citenum{tzenovBOOK} (pages 124--129), one can write the solution of the Hamilton's equations of motion as 
\begin{equation}
{\bf u} {\left( \theta \right)} = {\widehat{\cal R}} {\left( \theta; \theta_0 \right)} {\bf u}_0 + \int \limits_{\theta_0}^{\theta} {\rm d} \tau {\widehat{\cal R}} {\left( \theta; \theta_0 \right)} {\widehat{\cal R}}^{-1} {\left( \tau; \theta_0 \right)} {\bf F} {\left[ {\bf u} {\left( \tau \right)}; \tau \right]}. \label{SolHamil}
\end{equation}
Here ${\bf u} = {\left( X, P_x, Z, P_z \right)}^T$ is the state vector, and ${\bf F} = {\left( 0, - V_X, 0, - V_Z \right)}^T$, where the subscript below the potential function $V$ denotes partial derivative with respect to the variable indicated, and the superscript "$T$" implies matrix transposition. The matrix ${\widehat{\cal R}} {\left( \theta; \theta_0 \right)}$ of fundamental solutions to the unperturbed problem can be expressed as 
\begin{equation}
{\widehat{\cal R}} {\left( \theta; \theta_0 \right)} = 
\begin{pmatrix}
\cos \Delta \chi_x & \sin \Delta \chi_x & 0 & 0 \\ 
- \sin \Delta \chi_x & \cos \Delta \chi_x & 0 & 0 \\ 
0 & 0 & \cos \Delta \chi_z & \sin \Delta \chi_z \\
0 & 0 & - \sin \Delta \chi_z & \cos \Delta \chi_z 
\end{pmatrix}, \label{FunMatrix}
\end{equation}
where 
\begin{equation}
\Delta \chi_{x,z} {\left( \theta \right)} = \chi_{x,z} {\left( \theta \right)} - \chi_{x,z} {\left( \theta_0 \right)}, \label{PhaseAdv}
\end{equation}
is the phase advance between points $\theta$ and $\theta_0$. 

In what follows, we shall confine ourselves to the case of one degree of freedom, In particular, we shall analyse the dynamics in the horizontal with respect to the particle orbit direction. To be more specific in the choice of the nonlinear magnetic multipole element perturbing the linear machine structure, we consider the case of a single sextupole. Then, Eq. (\ref{SolHamil}) can be cast in a familiar already form [compare with Eq. (\ref{VoltFredVec})] as 
\begin{equation}
{\bf u} {\left( \theta \right)} + \int \limits_0^{\infty} {\rm d} \tau {\bf K} {\left( \theta; \tau \right)} u_1^2 {\left( \tau \right)} = {\bf v} {\left( \theta \right)}. \label{VoltHamil}
\end{equation}
Without loss of generality, the initial location along the machine circumference has been assumed to be $\theta_0 = 0$. In addition,  
\begin{equation}
{\bf K} {\left( \theta; \tau \right)} = {\cal S}_0 {\left( \tau \right)} \Theta {\left( \theta - \tau \right)}
\begin{pmatrix}
\sin {\left[ \chi {\left( \theta \right)} - \chi {\left( \tau \right)} \right]} \\ 
\cos {\left[ \chi {\left( \theta \right)} - \chi {\left( \tau \right)} \right]}
\end{pmatrix}, \label{Kernel}
\end{equation}
\begin{equation}
{\bf v} {\left( \theta \right)} = {\sqrt{2 J_0}} 
\begin{pmatrix}
\cos {\left[ \Delta \chi {\left( \theta \right)} + \alpha \right]} \\ 
- \sin {\left[ \Delta \chi {\left( \theta \right)} + \alpha \right]}
\end{pmatrix}, \label{VFunction}
\end{equation}
and the quantity ${\cal S}_0 {\left( \tau \right)}$ in Eq. (\ref{Kernel}) measures the sextupole strength 
\begin{equation}
{\cal S}_0 {\left( \theta \right)} = {\frac {\lambda_0 {\left( \theta \right)} \beta^{3/2} {\left( \theta \right)}} {2 R^2}}, \qquad \lambda_0 = {\frac {R^2} {B_z}} {\left( {\frac {\partial^2 B_z} {\partial x^2}} \right)}_{x,z=0}. \label{Sextupole}
\end{equation}
Moreover, $J_0$ is the initial value of the action variable, while $\alpha$ is the angle. Note also that the function $\Theta {\left( \theta - \tau \right)}$ in Eq. (\ref{Kernel}) is the standard Heaviside step function introduced to account for causality. In this way, the kernel ${\bf K} {\left( \theta; \tau \right)}$ itself is nothing but the Green's function for the classical harmonic oscillator. 

In the spirit of Eq. (\ref{BasSolVec}) representing the solution of the system of nonlinear integral equations (\ref{VoltFredVec}), of which kind through the identification 
\begin{equation}
{\bf G} {\left[ {\bf x}, {\bf u} \right]} = \int \limits_0^{\infty} {\rm d} \tau {\bf K} {\left( \theta; \tau \right)} u_1^2 {\left( \tau \right)}, \label{Identif} 
\end{equation}
is also Eq. (\ref{VoltHamil}), we expand the operator-exponent in the numerator as well as in the denominator of Eq. (\ref{BasSolVec}) in a formal Volterra series. The ratio of the latter two series can be expressed as 
\begin{equation}
{\bf u} {\left( \theta \right)} = {\dfrac {{\mathlarger{\sum\limits_{n=0}^{\infty}}} {\bf P}^{(n)} {\left( \theta \right)}} {{\mathlarger{\sum\limits_{n=0}^{\infty}}} Q^{(n)} {\left( \theta \right)}}}. \label{SolRatio} 
\end{equation}
Following the prescriptions and the explanatory notes in Section \ref{sec:funcgen} on how to use in practice the symbolic form of the solution expressed by Eq. (\ref{BasSolVec}), for the first several terms we obtain 
\begin{equation}
{\bf P}^{(0)} {\left( \theta \right)} = {\bf v}_0 {\left( \theta \right)}, \qquad \qquad Q^{(0)} = 1, \label{SolP0Q0} 
\end{equation}
\begin{equation}
{\bf P}^{(1)} {\left( \theta \right)} = - \int \limits_0^{\infty} {\rm d} \tau {\bf K} {\left( \theta; \tau \right)} v_1^2 {\left( \tau \right)}, \qquad \quad Q^{(1)} = 0, \label{SolP1Q1} 
\end{equation}
\begin{eqnarray}
Q^{(2)} = 2 \int \limits_0^{\infty} {\rm d} \sigma_1 \int \limits_0^{\infty} {\rm d} \sigma_2 K_1 {\left( \sigma_1; \sigma_2 \right)} \nonumber 
\\ 
\times K_1 {\left( \sigma_2; \sigma_1 \right)} v_1 {\left( \sigma_1 \right)} v_1 {\left( \sigma_2 \right)} \equiv 0, \label{SolQ2} 
\end{eqnarray}
\begin{eqnarray}
{\bf P}^{(2)} {\left( \theta \right)} = Q^{(2)} {\left( \theta \right)} {\bf v} {\left( \theta \right)} + 2 \int \limits_0^{\infty} {\rm d} \sigma \int \limits_0^{\infty} {\rm d} \tau {\bf K} {\left( \theta; \sigma \right)} \nonumber 
\\ 
\times K_1 {\left( \sigma; \tau \right)} v_1 {\left( \sigma \right)} v_1^2 {\left( \tau \right)}. \label{SolP2} 
\end{eqnarray}
It is essential to note that $Q^{(2)}$ in Eq. (\ref{SolQ2}) is identically equal to zero, thanks to the fact that $\Theta {\left( w \right)} \Theta {\left( - w \right)} = 0$ for arbitrary argument $w$. Also, by its very nature, the denominator is a simple number, since the exponential gamma-operator acts on unity instead of acting on a function dependent on the independent variable $\theta$. 
\begin{figure}
\begin{center} 
\includegraphics[width=8.0cm]{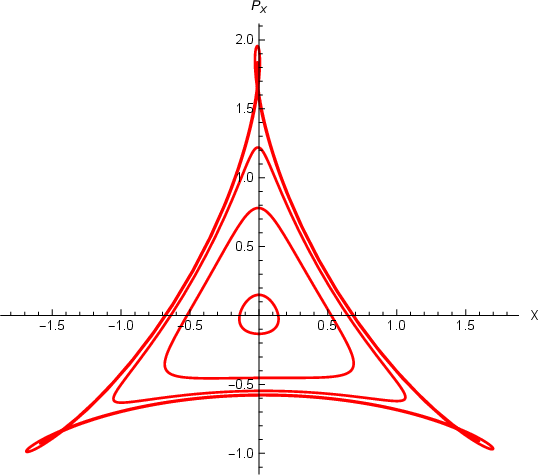}
\caption{\label{fig1:epsart} Phase portrait of betatron motion for 5500 turns obtained by taking into account only the first two terms ${\bf P}^{(0)} {\left( 2 \pi N \right)} + {\bf P}^{(1)} {\left( 2 \pi N \right)}$ in the numerator of Eq. (\ref{SolRatio}). The sextupole strength is ${\cal S} = 3.13577 \times 10^{-4}$, the fractional part of the betatron tune is $0.333333$, the initial value of the action variable for the inner-most curve is $J_0 = 0.01$, while for the outer curve is $J_0 = 0.8$.}
\end{center}
\end{figure}

In order to illustrate the results obtained by the method of formal series, we consider a single sextupole kick in the vicinity of $\theta = 0, 2 \pi, 4 \pi, \dots$. In thin lens approximation ${\cal S}_0 {\left( \theta \right)}$ in Eq. (\ref{Sextupole}) can be written as a sampling function (also known as Dirac comb function) 
\begin{equation}
{\cal S}_0 {\left( \theta \right)} = {\cal S} {\mathlarger{\sum\limits_{k=0}^{N}}} \delta {\left( \theta - 2k \pi \right)}, \qquad \quad {\cal S} = {\frac {L \lambda_0 \beta_0^{3/2}} {2 R^3}}, \label{Sampling} 
\end{equation}
where $N$ is the number of turns, and $L$ is the sextupole length. The integrals (\ref{SolP1Q1}) -- (\ref{SolP2}) become now finite sums with due account of the number of turns, which explicit form is given in Appendix \ref{sec:appendixB}. 
\begin{figure}
\begin{center} 
\includegraphics[width=8.0cm]{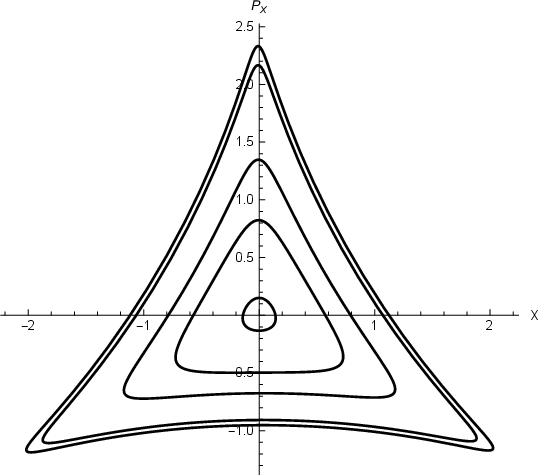}
\caption{\label{fig2:epsart} Phase portrait of betatron motion for 5500 turns obtained by taking into account all terms up to second order in both the numerator and the denominator of Eq. (\ref{SolRatio}). Similar to Fig. \ref{fig1:epsart} the sextupole strength is ${\cal S} = 3.13577 \times 10^{-4}$, the fractional part of the betatron tune is $0.333333$, the initial value of the action variable for the inner-most curve is $J_0 = 0.01$, while for the outer curve is $J_0 = 0.8$.}
\end{center}
\end{figure}

Figure \ref{fig1:epsart} shows the phase portrait of the nonlinear betatron oscillations close to the third order resonance $3 \nu = integer$ after 5500 turns along the accelerator circumference. Only the linear (with respect to the strength of the sextupole ${\cal S}$) terms in the general solution (\ref{SolRatio}) have been taken into account, which is equivalent to the standard Henon map \cite{Henon} widely used in nonlinear beam dynamics. For comparison, in Fig. \ref{fig2:epsart} the phase portrait is shown when using the full regularized solution given by Eq. (\ref{SolRatio}) up to second order with respect to the sextupole strength ${\cal S}$. The same parameters as in the case of Fig. \ref{fig1:epsart} have been considered. Most remarkably, even at sufficiently large values of the initial action variable $J_0$, the characteristic resonant triangle does not deform considerably, as in the case of the Henon map. Another interesting feature is that the maximum value extent at the vertices (unstable fixed points) of the characteristic resonant triangle with respect to both $X$ and $P_x$ increases a bit as $J_0$ grows, but the phase-space trajectories remain stable. This makes the method of formal series unique and distinctive among all other methods in nonlinear dynamics.

\section{\label{sec:hamiljacobi}Solution of the Hamilton-Jacobi Equation by the Method of Formal Series} 

Compared to the previous Section \ref{sec:solham}, there exist a much more elegant way to study the dynamics of nonlinear Hamiltonian systems, which we will consider now. Instead of the standard canonical variables in phase space, it is sometimes more convenient to use action angle variables. A significant number of problems in the physics of charged particle accelerators, plasma physics and celestial mechanics can be reduced to the analysis and more detailed study of the peculiarities of dynamical systems described by the Hamiltonian
\begin{equation}
{\cal H} = {\dot{\boldsymbol{\chi}}} {\left( \theta \right)} \cdot {\bf I} + V {\left( {\boldsymbol{\alpha}}, {\bf I}; \theta \right)}. \label{HamiActAng} 
\end{equation}
Here ${\bf I} = {\left( I_1, I_2, \dots, I_M \right)}$ and ${\boldsymbol{\alpha}} = {\left( \alpha_1, \alpha_2, \dots, \alpha_M \right)}$ are $M$-dimensional action angle variables, respectively. The quantities ${\boldsymbol{\chi}} = {\left( \chi_1, \chi_2, \dots, \chi_M \right)}$ represent the phase advances corresponding to the different degrees of freedom, and the dot above implies a derivative with respect to the "time" variable $\theta$. Furthermore, the perturbation potential $V {\left( {\boldsymbol{\alpha}}, {\bf I}; \theta \right)}$ is periodic in the angle variables, namely
\begin{eqnarray}
V {\left( {\boldsymbol{\alpha}}, {\bf I}; \theta \right)} = {\mathlarger{\sum\limits_{m_1,\dots,m_M}}} V_{m_1,\dots,m_M} {\left( {\bf I}; \theta \right)} \nonumber 
\\ 
\times \exp {\left( i \sum \limits_{k=1}^M m_k \alpha_k \right)}, \label{PertPot} 
\end{eqnarray}
where $V_{\bf 0} {\left( {\bf I}; \theta \right)} \neq 0$ in general. Following the standard procedure of canonical perturbation theory, we seek a canonical transformation given by the generating function
\begin{equation}
S {\left( {\boldsymbol{\alpha}}, {\bf J}; \theta \right)} = {\boldsymbol{\alpha}} \cdot {\bf J} + G {\left( {\boldsymbol{\alpha}}, {\bf J}; \theta \right)}, \label{GenerFunc} 
\end{equation}
such that the new action variables ${\bf J}$ are invariants of motion. This implies that the new Hamiltonian must depend on the new action variables only, and that is all we know about it yet. The equations defining the new canonical variables ${\left( {\bf a}, {\bf J} \right)}$ and the new Hamiltonian are
\begin{equation}
{\bf a} = {\boldsymbol{\alpha}} + {\boldsymbol{\partial}}  G {\left( {\boldsymbol{\alpha}}, {\bf J}; \theta \right)}, \qquad {\bf I} = {\bf J} + {\boldsymbol{\nabla}}  G {\left( {\boldsymbol{\alpha}}, {\bf J}; \theta \right)}, \label{NewAcAngVar} 
\end{equation}
\begin{equation}
{\overline{\cal H}} {\left( {\bf J}; \theta \right)} = {\dot{\boldsymbol{\chi}}} {\left( \theta \right)} \cdot {\bf J} + {\widehat{\cal P}}_{\bf M} V {\left( {\boldsymbol{\alpha}}, {\bf J} + {\boldsymbol{\nabla}}  G; \theta \right)}, \label{NewHamilt} 
\end{equation}
where for brevity ${\boldsymbol{\partial}} = {\left( \partial_{J_1}, \partial_{J_2}, \dots, \partial_{J_M} \right)}$ denotes the vector of partial derivatives with respect to the components of the action ${\bf J}$, while ${\boldsymbol{\nabla}} = {\left( \partial_{\alpha_1}, \partial_{\alpha_2}, \dots, \partial_{\alpha_M} \right)}$ is the vector of partial derivatives with respect to the components of the angle ${\boldsymbol{\alpha}}$. Moreover, the projection operator 
\begin{equation}
{\widehat{\cal P}}_{\bf M} \dots = {\frac {1} {{\left( 2 \pi \right)}^M}} \int \limits_0^{2 \pi} {\rm d}^M {\boldsymbol{\alpha}} \dots, \label{ProjOper} 
\end{equation}
indicates averaging over the angle variables. Therefore, the Hamilton-Jacobi equation can be written as 
\begin{equation}
\partial_{\theta} G + {\dot{\boldsymbol{\chi}}} \cdot {\boldsymbol{\nabla}} G + {\left( 1 - {\widehat{\cal P}}_{\bf M} \right)} V {\left( {\boldsymbol{\alpha}}, {\bf J} + {\boldsymbol{\nabla}}  G; \theta \right)} = 0. \label{HamilJacobi} 
\end{equation}
It can easily be verified by direct substitution and simple straightforward manipulation that the formal general solution of the Hamilton-Jacobi equation is 
\begin{eqnarray}
G {\left( {\boldsymbol{\alpha}}, {\bf J}; \theta \right)} + {\left( 1 - {\widehat{\cal P}}_{\bf M} \right)} 
\int \limits_0^{\infty} {\rm d} \tau {\widehat{\bf T}} {\left( \tau, \theta \right)} \nonumber 
\\ 
\times V {\left[ {\boldsymbol{\alpha}}, {\bf J} + {\boldsymbol{\nabla}}  G {\left( {\boldsymbol{\alpha}}, {\bf J}; \tau \right)}; \tau \right]} = G^{(0)} {\left( {\boldsymbol{\alpha}}, {\bf J}; \theta \right)}
, \label{SolHamiJac} 
\end{eqnarray}
where 
\begin{equation}
{\widehat{\bf T}} {\left( \tau, \theta \right)} = \Theta {\left( \theta - \tau \right)} \exp {\left\{ {\left[ {\boldsymbol{\chi}} {\left( \tau \right)} - {\boldsymbol{\chi}} {\left( \theta \right)} \right]} \cdot {\boldsymbol{\nabla}} \right\}}, \label{TransOper} 
\end{equation}
is a causal translation operator in the subspace of angle variables, and $G^{(0)}$ is a solution of the homogeneous part of the Hamilton-Jacobi equation 
\begin{equation}
\partial_{\theta} G^{(0)} + {\dot{\boldsymbol{\chi}}} \cdot {\boldsymbol{\nabla}}  G^{(0)} = 0. \nonumber 
\end{equation}

However, for the specific application of the method of formal series to the solution of partial differential equations, another alternative representation of the solution of the Hamilton-Jacobi equation (\ref{HamilJacobi}) will be more convenient in what follows. As can be easily verified, the Green's function ${\mathcal{G}} {\left( {\boldsymbol{\alpha}}, {\mathbf{A}}; \theta, \tau \right)}$ for the linear Hamilton-Jacobi operator $\partial_{\theta} G +{\dot{\boldsymbol{\chi}}} \cdot {\boldsymbol{\nabla}} G$ can be expressed as follows 
\begin{equation}
{\mathcal{G}} {\left( {\boldsymbol{\alpha}}, {\mathbf{A}}; \theta, \tau \right)} = \Theta {\left( \theta - \tau \right)} \delta {\left[ {\boldsymbol{\alpha}} - {\mathbf{A}} + {\boldsymbol{\chi}} {\left( \tau \right)} - {\boldsymbol{\chi}} {\left( \theta \right)} \right]}. \label{GreenFunct} 
\end{equation}
Thus, we obtain an alternative, as compared to that given by Eq. (\ref{SolHamiJac}), form of the general solution of the Hamilton-Jacobi equation 
\begin{eqnarray}
G {\left( {\boldsymbol{\alpha}}, {\bf J}; \theta \right)} + {\left( 1 - {\widehat{\cal P}}_{\bf M} \right)} 
\iint {\rm d} \tau {\rm d}^M {\mathbf{A}} {\mathcal{G}} {\left( {\boldsymbol{\alpha}}, {\mathbf{A}}; \theta, \tau \right)} \nonumber 
\\ 
\times V {\left[ {\mathbf{A}}, {\bf J} + {\boldsymbol{\nabla}}_{\mathbf{A}} G {\left( {\mathbf{A}}, {\bf J}; \tau \right)}; \tau \right]} = G^{(0)} {\left( {\boldsymbol{\alpha}}, {\bf J}; \theta \right)}
, \label{GreenHamJac} 
\end{eqnarray}
Differentiating both sides of the above Eq. (\ref{GreenHamJac}) with respect to the angles ${\boldsymbol{\alpha}}$, we transform it to a nonlinear Fredholm integral equation of the second kind \cite{WhitBOOK} 
\begin{eqnarray}
{\boldsymbol{\nabla}} G {\left( {\boldsymbol{\alpha}}, {\bf J}; \theta \right)} + \iint {\rm d} \tau {\rm d}^M {\mathbf{A}} {\boldsymbol{\mathcal{K}}} {\left( {\boldsymbol{\alpha}}, {\mathbf{A}}; \theta, \tau \right)} \nonumber 
\\ 
\times V {\left[ {\mathbf{A}}, {\bf J} + {\boldsymbol{\nabla}}_{\mathbf{A}} G {\left( {\mathbf{A}}, {\bf J}; \tau \right)}; \tau \right]} = {\boldsymbol{\nabla}} G^{(0)} {\left( {\boldsymbol{\alpha}}, {\bf J}; \theta \right)}
, \label{SolHamJacobi} 
\end{eqnarray}
for the unknown vector function ${\boldsymbol{\nabla}} G {\left( {\boldsymbol{\alpha}}, {\bf J}; \theta \right)}$, where 
\begin{equation}
{\boldsymbol{\mathcal{K}}} {\left( {\boldsymbol{\alpha}}, {\mathbf{A}}; \theta, \tau \right)} = {\boldsymbol{\nabla}} {\mathcal{G}} {\left( {\boldsymbol{\alpha}}, {\mathbf{A}}; \theta, \tau \right)}. \label{KernelHJ} 
\end{equation}

Identifying now 
\begin{equation}
{\bf u} {\left( {\boldsymbol{\alpha}}, \theta \right)} = {\boldsymbol{\nabla}} G {\left( {\boldsymbol{\alpha}}, {\bf J}; \theta \right)}, \qquad {\bf v} {\left( {\boldsymbol{\alpha}}, \theta \right)} = {\boldsymbol{\nabla}} G^{(0)} {\left( {\boldsymbol{\alpha}}, {\bf J}; \theta \right)}, \label{Identify1} 
\end{equation}
\begin{eqnarray}
{\bf G} {\left[ {\boldsymbol{\alpha}}, \theta, {\bf u} \right]} = \iint {\rm d} \tau {\rm d}^M {\mathbf{A}} {\boldsymbol{\mathcal{K}}} {\left( {\boldsymbol{\alpha}}, {\mathbf{A}}; \theta, \tau \right)} \nonumber 
\\ 
\times V {\left[ {\mathbf{A}}, {\bf J} + {\bf u} {\left( {\mathbf{A}}, \tau \right)}; \tau \right]}, \label{Identify2} 
\end{eqnarray}
it becomes clear that we managed to represent the formal solution of the Hamilton-Jacobi equation in the form of a nonlinear operator equation of the type (\ref{VoltFredVec}). Acting in a similar manner as in Section \ref{sec:solham}, we expand the operator-exponent in the numerator as well as in the denominator of Eq. (\ref{BasSolVec}) in a formal Volterra series and represent the latter as a ratio of two series in the form of Eq. (\ref{SolRatio}). In order to proceed further, we note that the initial condition $G^{(0)} {\left( {\boldsymbol{\alpha}}, {\bf J}; \theta \right)} \equiv 0$ automatically implies that ${\bf v} {\left( \theta \right)}$ should be set to zero at the end of all algebraic manipulations. Taking into account this fact, for the corresponding first several terms, we obtain 
\begin{widetext}
\begin{equation}
{\bf P}^{(0)} {\left( {\boldsymbol{\alpha}}, {\bf J}; \theta \right)} = 0, \qquad \qquad Q^{(0)} {\left( {\bf J} \right)} = 1, \label{SolHJP0Q0} 
\end{equation}
\begin{equation}
{\bf P}^{(1)} {\left( {\boldsymbol{\alpha}}, {\bf J}; \theta \right)} = - \int {\rm d} \tau \Theta {\left( \theta - \tau \right)} {\boldsymbol{\nabla}} V {\left( {\boldsymbol{\alpha}} + {\boldsymbol{\chi}} {\left( \tau \right)} - {\boldsymbol{\chi}} {\left( \theta \right)}, {\bf J}; \tau \right)}, \label{SolHJP1Q1} 
\end{equation}
\begin{equation}
{\left. Q^{(1)} {\left( {\bf J} \right)} = - \int {\rm d} \sigma {\left( {\boldsymbol{\partial}} \cdot {\boldsymbol{\nabla}} \right)} V {\left( {\boldsymbol{\alpha}}, {\bf J}; \sigma \right)} \right|}_{{\boldsymbol{\alpha}} = 0}, \label{SolHJP1Q12} 
\end{equation}
\begin{equation}
Q^{(2)} {\left( {\bf J} \right)} = {\frac {1} {2}} Q^{(1) {\bf 2}} {\left( {\bf J} \right)} + {\left. \iint {\rm d} \sigma {\rm d} \tau \Theta {\left( \sigma - \tau \right)} {\boldsymbol{\nabla}}_k {\left[ {\boldsymbol{\partial}}_k {\boldsymbol{\partial}}_m V {\left( {\boldsymbol{\alpha}}, {\mathbf{J}}; \sigma \right)} {\boldsymbol{\nabla}}_m V {\left( {\boldsymbol{\alpha}} + {\boldsymbol{\chi}} {\left( \tau \right)} - {\boldsymbol{\chi}} {\left( \sigma \right)}, {\mathbf{J}}; \tau \right)} \right]} \right|}_{{\boldsymbol{\alpha}} = 0}, \label{SolHJQ2} 
\end{equation}
\begin{equation}
{\bf P}^{(2)} {\left( {\boldsymbol{\alpha}}, {\bf J}; \theta \right)} = Q^{(1)} {\left( {\bf J} \right)} {\bf P}^{(1)} {\left( {\boldsymbol{\alpha}}, {\bf J}; \theta \right)} - \int {\rm d} \tau \Theta {\left( \theta - \tau \right)} 
{\boldsymbol{\nabla}} {\left[ {\boldsymbol{\partial}} V {\left( {\boldsymbol{\alpha}} + {\boldsymbol{\chi}} {\left( \tau \right)} - {\boldsymbol{\chi}} {\left( \theta \right)}, {\mathbf{J}}; \tau \right)} \cdot {\mathbf{P}}^{(1)} {\left( {\boldsymbol{\alpha}} + {\boldsymbol{\chi}} {\left( \tau \right)} - {\boldsymbol{\chi}} {\left( \theta \right)}, {\mathbf{J}}; \tau \right)} \right]}. \label{SolHJP2} 
\end{equation}
\end{widetext}
Summation over repeated indices in the expression (\ref{SolHJQ2}) for $Q^{(2)}$ is implied. 
\begin{figure}
\begin{center} 
\includegraphics[width=8.0cm]{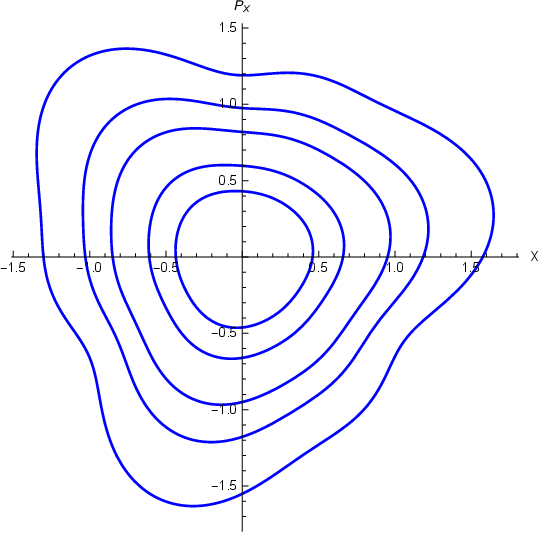}
\caption{\label{fig3:epsart} Phase portrait of betatron motion for 5000 turns obtained by taking into account all terms up to second order in both the numerator and the denominator of Eq. (\ref{SolRatio}). For the corresponding terms Eqs. (\ref{SolHJP0Q0}) -- (\ref{SolHJP2}) have been taken into account. The sextupole strength is ${\cal S} = 3.13577 \times 10^{-4}$, the fractional part of the betatron tune is $0.3332809$, the initial value of the action variable for the inner-most curve is $J_0 = 0.1$, while for the outer curve is $J_0 = 1.0$.}
\end{center}
\end{figure}

The frequency of nonlinear oscillations can be determined according to the expression
\begin{eqnarray}
\omega_n {\left( {\bf J} \right)} = \nu_n + {\widehat{\cal P}}_{\bf M} {\left< {\left[ \delta_{n k} + \partial_n \nabla_k G {\left( {\boldsymbol{\alpha}}, {\bf J}; \theta \right)} \right]} \right.} \nonumber 
\\ 
{\left. \times \partial_k V {\left( {\boldsymbol{\alpha}}, {\bf J} + {\boldsymbol{\nabla}} G; \theta \right)} \right>}_{\theta}, \label{Frequency} 
\end{eqnarray}
where 
\begin{equation}
{\left< \dots \right>}_{\theta} = {\frac {1} {2 \pi}} \int \limits_0^{2 \pi} {\rm d} \theta \dots, \label{Average} 
\end{equation}
denotes averaging over one turn and summation over repeated indices is implied. Thus, the nonlinear tune shift $\Delta \nu_n {\left( {\bf J} \right)} = \omega_n {\left( {\bf J} \right)} - \nu_n$ can be immediately calculated. Once the generator ${\boldsymbol{\nabla}} G$ has been determined, it is straightforward to reproduce the phase portrait of the system scanned at subsequent times. This can be done by utilizing Eq. (\ref{NewAcAngVar}) for different values of the action invariant $\bf J$. 

\subsection{\label{sec:sextupole}Sextupole} 

To illustrate the results obtained in this Section, let us consider again a single sextupole kick in the vicinity of $\theta = 0, 2 \pi, 4 \pi, \dots$. In thin lens approximation, we use the representation (\ref{Sampling}) as before.  
\begin{figure}
\begin{center} 
\includegraphics[width=8.0cm]{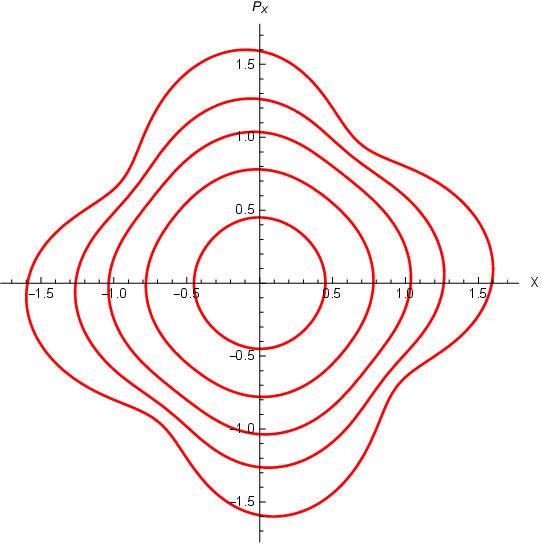}
\caption{\label{fig4:epsart} Phase portrait of betatron motion for 5000 turns obtained by taking into account all terms up to second order in both the numerator and the denominator of Eq. (\ref{SolRatio}). For the corresponding terms Eqs. (\ref{SolHJP0Q0}) -- (\ref{SolHJP2}) have been taken into account. The octupole strength is ${\cal O} = 1.11436 \times 10^{-4}$, the fractional part of the betatron tune is $0.24997$, the initial value of the action variable for the inner-most curve is $J_0 = 0.1$, while for the outer curve is $J_0 = 0.98$.}
\end{center}
\end{figure}
Figure \ref{fig3:epsart} shows the phase portrait of the nonlinear betatron oscillations close to the third order resonance $3 \nu = integer$ after 5000 turns along the accelerator circumference. The complete regularized solution given by Eq. (\ref{SolRatio}) with due account of Eqs. (\ref{SolHJP0Q0}) -- (\ref{SolHJP2}) up to second order with respect to the sextupole strength ${\cal S}$ has been taken into account. The same values of the basic parameters as in the case of Figs. \ref{fig1:epsart} and \ref{fig2:epsart} have been considered.

\subsection{\label{sec:octupole}Octupole} 

As a second illustrative example, we shall consider the case of a single octupole kick in the vicinity of $\theta = 0, 2 \pi, 4 \pi, \dots$. In thin lens approximation the octupole strength ${\cal O}_0 {\left( \theta \right)}$ can be written as 
\begin{equation}
{\cal O}_0 {\left( \theta \right)} = {\cal O} {\mathlarger{\sum\limits_{k=0}^{N}}} \delta {\left( \theta - 2k \pi \right)}, \qquad \quad {\cal O} = {\frac {L_o \mu_0 \beta_0^2} {6 R^4}}, \label{SamplingOct} 
\end{equation}
where $N$ is the number of turns, and $L_o$ is the octupole length. Here 
\begin{equation}
\mu_0 = {\frac {R^3} {B_z}} {\left( {\frac {\partial^3 B_z} {\partial x^3}} \right)}_{x=z=0}. \label{Mu0} 
\end{equation}
\begin{figure}
\begin{center} 
\includegraphics[width=8.0cm]{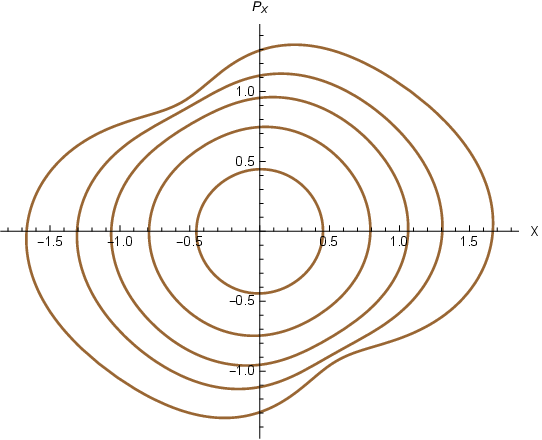}
\caption{\label{fig5:epsart} Phase portrait of betatron motion for 5000 turns obtained by taking into account all terms up to second order in both the numerator and the denominator of Eq. (\ref{SolRatio}). For the corresponding terms Eqs. (\ref{SolHJP0Q0}) -- (\ref{SolHJP2}) have been taken into account. The octupole strength is ${\cal O} = 1.07626 \times 10^{-4}$, the fractional part of the betatron tune is $0.499933$, the initial value of the action variable for the inner-most curve is $J_0 = 0.1$, while for the outer curve is $J_0 = 0.98$.}
\end{center}
\end{figure}

Figure \ref{fig4:epsart} shows the phase portrait of the nonlinear betatron oscillations close to the fourth order resonance $4 \nu = integer$ after 5000 turns along the accelerator circumference. The complete regularized solution given by Eq. (\ref{SolRatio}) with due account of Eqs. (\ref{SolHJP0Q0}) -- (\ref{SolHJP2}) up to second order with respect to the octupole strength ${\cal O}$ has been taken into account. Finally, Fig. \ref{fig5:epsart} shows the phase portrait of the nonlinear betatron oscillations close to the parametric resonance $2 \nu = integer$ after 5000 turns along the accelerator circumference. 

\section{\label{sec:concluding}Concluding Remarks} 

As a result of the study conducted here, it is shown that the method of formal series can be successfully applied to solve a number of problems in the physics of accelerators and charged particle beams. The range of problems studied here is not limited to the dynamics of charged particles in accelerators, but is quite general, which makes the obtained results applicable to nonlinear dynamics in general. The solution of Hamilton's equations describing the particle dynamics as well as the Hamilton-Jacobi equation can be represented as a ratio of two formal Volterra series depending on the perturbation parameter. It is written in a compact and elegant form and can be handled explicitly to a very high order. Furthermore, the method can be numerically implemented in a straightforward manner to obtain both analytical, as well as numerical results that would provide a direct way of assessing the effect of high order resonances. It is worthwhile to note that if the denominator of the fraction on the right-hand-side of Eq. (\ref{SolRatio}) is re-expanded in the formal perturbation parameter and the multiplication performed, one ends up with an expression that can be obtained by a direct perturbation expansion of the solution to Eq. (\ref{SolHamJacobi}).

The method of formal series offers a new approach to tackling a number of problems in accelerator physics, either by direct solution of the equations of motion, or by solution of the Hamiltonian-Jacobi equation. Remarkable is the comparison of Figure \ref{fig1:epsart} with Figure \ref{fig2:epsart}, where the smoothing effect on the invariant curves stands out. 

In addition to the applicability for the study of regular motion, the method of formal series seems a very promising tool for analysis and detailed investigation of the transition to chaos, especially in the case of systems with two or more degrees of freedom. Let us remind that according to the construction adopted here, the solution of the Hamilton-Jacobi equation describes the manifold of invariant curves. Naturally, the question arises, when and at what values of the input parameters do these invariant curves begin to break down. To answer this question, it is necessary that the new Hamiltonian be chosen differently and the classical modification of the scattering amplitude be applied for the appropriate purpose \cite{Rowe,Williams}. An idea of how this can be achieved, which, of course, needs to be further worked out, is presented in Appendix \ref{sec:appendixC}. Then, the denominator in Eq. (\ref{SolRatio}) will provide detailed information concerning the pole distribution for certain values of the nonlinear invariant and the strength of the nonlinearity, considering the unperturbed betatron tune as an input parameter. This suggests a possible analogy with Regge poles \cite{Regge} extensively used in quantum mechanics. Further refinement of the method described here could be the development of tools similar to the diagram technique widely used in quantum field theory in order to handle higher-order terms in the formal series expansion of the ${\widehat{\boldsymbol{\Gamma}}}$-exponents. In Appendix \ref{sec:appendixC}, the idea of representing the influence of a concentrated nonlinearity on the dynamics of a linear system as a classical scattering amplitude is vaguely implied. To illustrate the idea, the study has been conducted in the so-called smooth approximation. The smooth approximation is not a significant limitation, as the approach can be developed in the most general case. In our opinion, this approach is very promising to be pursued and developed further in more detail. Thus, the classical scattering amplitude embedded in the generator $G {\left( X, p; \theta \right)}$ can be studied analytically both by the methods of the standard canonical perturbation theory and by the method of formal series. We intend to tackle all of these in the near future.

We believe that the method of formal series will prove itself useful in lattice design, as well as in dynamic aperture studies. It could be incorporated in optimization strategies for tune shift correction, the distortion of invariant curves, and elimination of other undesirable effects. 

\begin{acknowledgments}

Fruitful discussions on topics touched upon in the present article with Drs. Zhichu Chen and Jianhui Chen are gratefully acknowledged.

\end{acknowledgments}

\appendix

\section{\label{sec:appendixA}Proof of the Basic Identity in Eq. (\ref{BasicIdentity})}

For the sake of simplicity and visibility, the proof of the identity in Eq. (\ref{BasicIdentity}) will be performed here in the one-dimensional case. Indeed, the generalization in the $N$-dimensional case is more cumbersome, but in principle, it can be done without difficulty. The proof consists in a sequence of detailed transformations and reorganizations of the sums entering, in the main lines, into the left-hand-side of the identity in question. In one-dimension, the left-hand-side of Eq. (\ref{BasicIdentity}) can be written as 
\begin{equation}
{\widehat{\boldsymbol{\Gamma}}} \exp {\left[ - \epsilon \partial_{\xi} g {\left( \xi \right)} \right]} \bullet f {\left[ \xi + \epsilon g {\left( \xi \right)} \right]} \nonumber 
\end{equation}
\begin{equation}
= {\mathlarger{\sum\limits_{m,n=0}^{\infty}}} {\frac {{\left( -1 \right)}^m \epsilon^{m+n}} {m! \; n!}} \partial_{\xi}^m {\left( g^{m+n} \partial_{\xi}^n f \right)} \nonumber 
\end{equation}
\begin{equation}
= {\mathlarger{\sum\limits_{p=0}^{\infty}}} {\left( -1 \right)}^p \epsilon^p {\mathlarger{\sum\limits_{n=0}^p}} {\frac {{\left( -1 \right)}^n} {n! \; (p-n)!}} \partial_{\xi}^{p-n} {\left( g^p \partial_{\xi}^n f \right)}, \label{LHSIdent}
\end{equation}
where $\partial_{\xi}$ denotes differentiation with respect to $\xi$. In passing from the second to the third row in the above equation, a change $m + n = p$ of the summation indices has been carried out. 

It suffices to prove that the second sum (see the third row) on the right-hand-side of Eq. (\ref{LHSIdent}) can be converted to the following closed form 
\begin{equation}
{\mathlarger{\sum\limits_{n=0}^p}} {\frac {{\left( -1 \right)}^n} {n! \; (p-n)!}} \partial_{\xi}^{p-n} {\left( g^p \partial_{\xi}^n f \right)} = {\frac {f} {p!}} \partial_{\xi}^p {\left( g^p \right)}. \label{Suffice}
\end{equation}
To prove this, let us write in detail in an explicit form the sum on the left-hand-side of Eq. (\ref{Suffice}). We have 
\begin{widetext}
\begin{eqnarray}
{\frac {1} {p!}} {\left( g^p f \right)}^{(p)} - {\frac {1} {(p-1)!}} {\left( g^p f^{\prime} \right)}^{(p-1)} + {\frac {1} {2! \; (p-2)!}} {\left( g^p f^{\prime \prime} \right)}^{(p-2)} - {\frac {1} {3! \; (p-3)!}} {\left( g^p f^{\prime \prime \prime} \right)}^{(p-3)} \nonumber
\\ 
+ \dots + {\frac {(-1)^{p-1}} {(p-1)!}} {\left( g^p f^{(p-1)} \right)}^{\prime} + {\frac {(-1)^p} {p!}}  g^p f^{(p)}
 \label{Explicit1}
\end{eqnarray}
\begin{eqnarray}
= && {\mathlarger{\sum\limits_{k=0}^p}} {\frac {1} {k! \; (p-k)!}} {\left( g^p \right)}^{(p-k)} f^{(k)} - {\mathlarger{\sum\limits_{k=0}^{p-1}}} {\frac {1} {k! \; (p-k-1)!}} {\left( g^p \right)}^{(p-k-1)} f^{(k+1)} + {\frac {1} {2!}} {\mathlarger{\sum\limits_{k=0}^{p-2}}} {\frac {1} {k! \; (p-k-2)!}} {\left( g^p \right)}^{(p-k-2)} f^{(k+2)} \nonumber \\ 
&& - {\frac {1} {3!}} {\mathlarger{\sum\limits_{k=0}^{p-3}}} {\frac {1} {k! \; (p-k-3)!}} {\left( g^p \right)}^{(p-k-3)} f^{(k+3)} + \dots + {\frac {(-1)^{p-1}} {(p-1)!}} {\left[ g^p f^{(p)} + {\left( g^p \right)}^{\prime} f^{(p-1)} \right]} + {\frac {(-1)^{p}} {p!}} g^p f^{(p)}, \label{Explicit}
\end{eqnarray}
\end{widetext}
where for brevity the notation $\partial_{\xi}^n (\dots) = (\dots)^{(n)}$ has been used. Looking closely at the right-hand-side of the above equality, one can observe that the coefficient multiplying the combination of derivatives of the kind ${\left( g^p \right)}^{(p-s)} f^{(s)}$, where $s = 1, 2, \dots, p$, is expressed as 
\begin{equation}
{\frac {1} {(p-s)!}} {\mathlarger{\sum\limits_{k=0}^s}} {\frac {{\left( -1 \right)}^k} {k! \; (s-k)!}} = 0. \label{BinomIdent}
\end{equation}
The equality above is a simple binomial identity, directly following from 
\begin{eqnarray}
0 = {\left( x - x \right)}^n = {\mathlarger{\sum\limits_{k=0}^n}} {\binom {n} {k}} (-1)^k x^{n-k} x^k \nonumber 
\\ 
= n! \; x^n {\mathlarger{\sum\limits_{k=0}^n}} {\frac {{\left( -1 \right)}^k} {k! \; (n-k)!}}, \label{BinomProof}
\end{eqnarray}
valid for arbitrary $x$. Thus, all terms on the left-hand-side of Eq. (\ref{Suffice}) containing derivatives of the function $f {\left( \xi \right)}$ cancel away, so that only the term on the right-hand-side survives. This completes the proof of the identity in Eq. (\ref{BasicIdentity}). 

\section{\label{sec:appendixB}Explicit Form of the Sums in the Thin Lens Approximation}

Substitution of Eq. (\ref{Sampling}) into Eqs. (\ref{SolP1Q1}) -- (\ref{SolP2}) gives 
\begin{eqnarray}
{\bf P}^{(1)} {\left( 2 \pi N \right)} = - 2 J_0 {\cal S} {\mathlarger{\sum\limits_{m=0}^N}} 
\begin{pmatrix}
\sin m \omega \\ 
\cos m \omega 
\end{pmatrix} \nonumber 
\\ 
\times \cos^2 {\left[ {\left( m - N \right)} \omega - \alpha \right]}, \label{SumP1} 
\end{eqnarray}
\begin{eqnarray}
&& {\bf P}^{(2)} {\left( 2 \pi N \right)} = 2 {\left( 2 J_0 \right)}^{3/2} {\cal S}^2 {\mathlarger{\sum\limits_{m=0}^N}} {\mathlarger{\sum\limits_{n=0}^m}} 
\begin{pmatrix}
\sin {\left[ {\left( N - m \right)} \omega \right]} \\ 
\cos {\left[ {\left( N - m \right)} \omega \right]} 
\end{pmatrix} \nonumber 
\\ 
&& \times \sin {\left[ {\left( m - n \right)} \omega \right]} \cos {\left( m \omega + \alpha \right)} \cos^2 {\left( n \omega + \alpha \right)}. \label{SumP2} 
\end{eqnarray}
Here $\omega = 2 \pi \nu$, where $\nu$ is the unperturbed betatron tune in the horizontal direction. The above sums can be calculated in a closed form. For example, by using simple trigonometric identities, the expression under the sum for $P_1^{(1)} {\left( 2 \pi N \right)}$ should be manipulated as follows 
\begin{eqnarray}
\sin {\left( m \omega \right)} \cos^2 {\left[ m \omega - {\left( N \omega + \alpha \right)} \right]} \nonumber 
\\ 
= {\frac {1} {4}} \cos 2 {\left( N \omega + \alpha \right)} \sin {\left( 3m \omega \right)} \nonumber 
\\ 
+ {\frac {1} {4}} {\left[ 2 - \cos 2 {\left( N \omega + \alpha \right)} \right]} \sin m \omega \nonumber 
\\ 
+ {\frac {1} {4}} \sin 2 {\left( N \omega + \alpha \right)} {\left[ \cos m \omega - \cos {\left( 3m \omega \right)} \right]}
, \label{Manipulate}
\end{eqnarray}
and then the expression \cite{PrudBOOK}
\begin{eqnarray}
{\mathlarger{\sum\limits_{k=0}^N}} 
\begin{pmatrix}
\sin {\left( k z + \varphi \right)} \\ 
\cos {\left( k z + \varphi \right)} 
\end{pmatrix} = {\rm cosec} {\left( {\frac {z} {2}} \right)} \nonumber 
\\ 
\times \sin {\frac {{\left( N + 1 \right)} z} {2}} 
\begin{pmatrix}
\sin {\left[ {\left( N z / 2 \right)} + \varphi \right]} \\ 
\cos {\left[ {\left( N z / 2 \right)} + \varphi \right]} 
\end{pmatrix}
, \label{SumClosed}
\end{eqnarray}
can be used to cast the sum in Eq. (\ref{SumP1}) in a closed form. Similarly, all trigonometric sums in the expressions above can be converted to closed form.

In the same way, all the products of trigonometric functions and their powers entering Eqs. (\ref{SolHJP0Q0}) -- (\ref{SolHJP2}) for the case of a sextupole, as well as for an octupole can be represented as sums of trigonometric functions of multiple arguments and the latter transformed into closed form. 

\begin{figure}
\begin{center} 
\includegraphics[width=8.0cm]{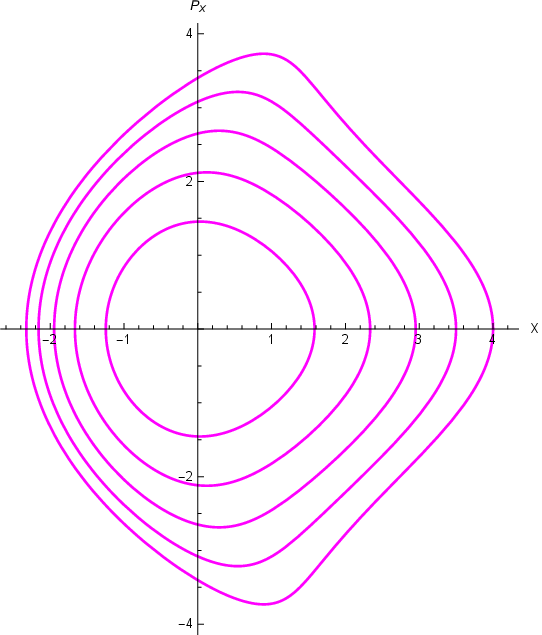}
\caption{\label{fig6:epsart} Phase portrait of the scattering orbits for 11500 turns obtained by taking into account Eq. (\ref{SolutFirstOrd}). The sextupole strength is ${\cal S} = 3.13577 \times 10^{-4}$, the fractional part of the betatron tune is $0.333333$, the initial value of the action variable for the inner-most curve is $J_0 = 1.0$, while for the outer curve is $J_0 = 5.0$.}
\end{center}
\end{figure}

\section{\label{sec:appendixC}Classical Scattering Amplitude}

For simplicity, let us consider again a single degree of freedom, i.e. the betatron motion in the horizontal ${\left( X, P_x \right)}$ plane in phase space in the presence of an isolated sextupole perturbing the linear machine lattice. In order to formulate the scattering problem from the sextupole potential, it is necessary to specify the Hamiltonian of the incoming and outgoing scattering channels. The initial (old) Hamiltonian is set by Eq. (\ref{HamilNorm}), while the Hamiltonian of the outgoing channel, which for us will be the new Hamiltonian, is 
\begin{equation}
{\overline{H}} {\left( x, p; \theta \right)} = {\frac {{\dot{\chi}}} {2}} {\left( p^2 + x^2 \right)}. \label{NewHamil} 
\end{equation}
Then, the Hamilton-Jacobi equation for the generating function $F {\left( X, p; \theta \right)}$ depending on the old coordinate $X$ and the new momentum $p$ reads as 
\begin{equation}
F_{\theta} + {\frac {{\dot{\chi}}} {2}} {\left( F_X^2 + X^2 \right)} + {\frac {\lambda_0 \beta^{3/2}} {6 R^2}} X^3 = {\frac {{\dot{\chi}}} {2}} {\left( p^2 + F_p^2 \right)}, \label{HamJacScatt} 
\end{equation}
where $F_w$ for $w = {\left( \theta, X, p \right)}$ denotes partial derivative with respect to the variable indicated. Using again the well-known representation 
\begin{equation}
F {\left( X, p; \theta \right)} = X p + G {\left( X, p; \theta \right)}, \nonumber 
\end{equation}
we rewrite Eq. (\ref{HamJacScatt}) as 
\begin{equation}
G_{\theta} + {\dot{\chi}} {\left( p G_X - X G_p \right)} + {\frac {{\dot{\chi}}} {2}} {\left( G_X^2 - G_p^2 \right)} + {\frac {\lambda_0 \beta^{3/2}} {6 R^2}} X^3 = 0, \label{HamJacScat} 
\end{equation}
To acquire an idea of the result of the solution of Eq. (\ref{HamJacScat}), let us perform an average over $N$ turns in smooth focusing approximation. Note that this approximation is not essential, an analytical solution can be found in a straightforward manner in the non-autonomous (explicit $\theta$-dependence) case as well. In the first order with respect to the strength of the sextupole, Eq. (\ref{HamJacScat}) can be written in the form 
\begin{equation}
p G_X^{(1)} - X G_p^{(1)} + {\frac {N {\cal S}} {3 \nu}} X^3 = 0. \label{HamJacScat1O} 
\end{equation}
Its solution reads as 
\begin{equation}
G^{(1)} {\left( X, p \right)} = {\frac {N {\cal S}} {3 \nu}} {\left( X^2 p + {\frac {2} {3}} p^3 \right)}. \label{SolutFirstOrd} 
\end{equation}
One can continue further and reconstruct the generator $G {\left( X, p \right)}$ into subsequent higher orders of magnitude. Figure \ref{fig6:epsart} shows the family of scattering orbits near a third order resonance taking into account the same parameters used in the main body of the article.






\bibliographystyle{unsrt}
\bibliography{formal}

\end{document}